\documentclass[twocolumn,aps,prl,showpacs]{revtex4}
\usepackage{amsmath}

\input epsf

\begin{document}

\title{Topological defects and bulk melting of hexagonal ice}
\author{Davide Donadio}
\affiliation{Computational Science, Department of Chemistry and Applied Biosciences, ETH
Z\"urich, USI Campus, Via Giuseppe Buffi 13, CH-6900 Lugano, Switzerland.}
\author{Paolo Raiteri}
\affiliation{Computational Science, Department of Chemistry and Applied Biosciences, ETH
Z\"urich, USI Campus, Via Giuseppe Buffi 13, CH-6900 Lugano, Switzerland.}
\author{Michele Parrinello}
\affiliation{Computational Science, Department of Chemistry and Applied Biosciences, ETH
Z\"urich, USI Campus, Via Giuseppe Buffi 13, CH-6900 Lugano, Switzerland.}

\begin{abstract}
We use classical molecular dynamics combined with the recently developed
metadynamics method $[$A. Laio and M. Parrinello, Procs. Natl. Acad. Sci.
USA \textbf{99}, 20 (2002)$]$ to study the process of bulk melting in
hexagonal ice. Our simulations show that bulk melting is mediated by the
formation of topological defects which preserve the coordination of the
tetrahedral network. Such defects cluster to form a defective region
involving about 50 molecules with a surprisingly long life-time. The
subsequent formation of coordination defects triggers the transition to the
liquid state.
\end{abstract}

\pacs{64.70.Dv,64.60.My,61.72.Ji}
\maketitle

The melting of ice is a process of obvious universal relevance. Under
ordinary circumstances melting is nucleated at surfaces, but situations
could be envisaged in which the effect of the surfaces is inhibited and
melting becomes a bulk effect \cite{mishima84}. Studying melting under these
circumstances can give very precious information on the nature of the
potential energy surface (PES) and on the transition to the disordered
state. One could in fact imagine that melting takes place either as a sudden
collapse of the lattice or through the creation and successive catastrophic
multiplication of defects \cite{fecht92}.

In principle molecular dynamics (MD) is an ideal tool for studying bulk
melting, since the imposed periodic boundary conditions eliminate surface
effects. Unfortunately the time scale over which melting occurs is too long
for present-day computational resources. This means that melting can be
observed only by super-heating the system to the point of inducing a sudden
lattice instability. Very recently we have developed the metadynamics
method, which allows long time scale phenomena to be studied \cite%
{laio02,marcella,micheletti}. Using metadynamics we are able to study the
melting of ice near the melting temperature of the adopted empirical model ($%
\sim $270 K) and we find that bulk melting is mediated by the formation of
topological defects, named \textquotedblleft 5+7\textquotedblright\ and
characterized in a recent work \cite{grishina04}. However, in contrast to
the picture of a sudden multiplication of these defects we find that before
melting ice goes through a metastable state where a defective region of
about 50 molecules is surrounded by an otherwise perfect lattice.

In order to overcome the time-scale problem, we exploit the extended
Lagrangian implementation of metadynamics \cite{marcella}. The method is
based on the construction of a coarse-grained non-Markovian dynamics in the
space defined by a few collective coordinates. The dynamics is biased by a
history-dependent potential term that, in time, fills the free energy
minima, allowing an efficient exploration and an accurate determination of
the free energy surface (FES). As in ref. \cite{marcella}, we introduce an
extended Hamiltonian that couples the collective variables $s_{\alpha
}\left( \mathbf{r}\right) $ to a set of additional dynamic variables $%
s_{\alpha }$: 
\begin{eqnarray}
H &=&H_{0}+\sum_{\alpha }\left[ \frac{1}{2}M_{\alpha }\left( \frac{%
ds_{\alpha }}{dt}\right) ^{2}+\frac{1}{2}k_{\alpha }\left( s_{\alpha
}-s_{\alpha }\left( \mathbf{r}\right) \right) ^{2}\right]   \notag \\
&&+V(s_{\alpha },t)  \label{h1}
\end{eqnarray}%
where $\mathbf{r}$ are the microscopic coordinates of the system and $H_{0}$
the unperturbed Hamiltonian. The masses $M_{\alpha }$ and the coupling
constants $k_{\alpha }$ should be chosen so as to achieve an efficient
coupling between the microscopic system and the collective variables, which
is obtained when the frequencies determined by $M_{\alpha }$ and $k_{\alpha }
$ are of the same order of magnitude as the characteristic frequencies of
the microscopic system. The history-dependent potential is defined as: 
\begin{equation}
V\left( s_{\alpha },t\right) =\sum_{\substack{ t_{i}=\left( \Delta t,2\Delta
t,\dots \right)  \\ t_{i}<t}}w\exp \left( -\frac{\left\Vert s_{\alpha
}-s_{\alpha }\left( t_{i}\right) \right\Vert ^{2}}{2\delta s^{2}}\right) 
\label{h2}
\end{equation}%
where the time interval $\Delta t$ between the placement of two successive
Gaussians, the Gaussian width $\delta s$ and the Gaussian height $w$ are
free parameters that affect the efficiency and the accuracy of the algorithm 
\cite{micheletti}. This method has already been applied successfully to the
study of rare events occurring in biological systems \cite{matteo}, chemical
reactions \cite{iannuzzi04} and phase transitions \cite{roman}.

Two different sets of collective variables were used in the study of the
melting transition. Common to both sets is the use of the potential energy,
which is a relevant order parameter for the melting as it undergoes a
sizable change when the phase transition takes place. Moreover it is
suitable for use as a collective variable in the metadynamics scheme, as it
is an explicit function of the microscopic configuration of the system \cite%
{micheletti}. The potential energy was supplemented by coordinates measuring
the intermediate order. In one case we exploited the Steinhardt \cite{q6}
order parameter Q$_{6}$, which has already been used successfully to
simulate the nucleation of ice I$_{h}$ in liquid water \cite{trout03}, and
in the other the number of 5- and 6-membered rings. In either case we find
similar results. Here we discuss only results obtained on the basis of the
second choice.

In order to compute the ring statistics we use the \emph{shortest path ring}
definition \cite{king,yuan02}. This is obtained by considering a H$_{2}$O
molecule and two of its nearest neighbors (n.n.) and finding the shortest
path passing through these three molecules. This criterion fails when
counting large primitive rings \cite{yuan02}, but these are of little
importance for our study. For use in the metadynamics it is necessary to
turn this definition into a continuous differentiable function of the atomic
coordinates $\mathcal{N}_{n}(\{\mathbf{r}_{i}\})$ which gives the number of $%
n$-membered rings. 
We define the hydrogen bond between water molecules $i$ and $j$ by a
function $f_{ij}$ of the atomic coordinates, which is 1 when the two
molecules are bonded and otherwise tends smoothly to zero \cite%
{swfun,paolowater}. For each H$_{2}$O molecule we consider the triplets $T$
formed by itself and the pairs of its neighbors and we compute the products $%
F_{T}^{I}=\prod_{i,j\subset I}f_{ij}$ of the hydrogen bond functions in the
closed paths $I$ containing $T$. The shortest path ring containing $T$ is
selected by a function $G_{T}$ defined as: 
\begin{equation}
G_{T}=\lambda /ln\sum_{I\supset T}e^{\lambda {F_{T}}^{I}/L_{I}}
\end{equation}%
where $L_{I}$ is the topological length of the path $I$. In the limit of
large $\lambda $, $G_{T}$ tends to the length of the shortest path ring. The
total number of $n$-membered rings is then given by a sum of rational
functions: 
\begin{equation}
\mathcal{N}_{n}(\{\mathbf{r}_{i}\})=\sum_{T}\frac{1-\Big(\frac{n-G_{T}}{%
\sigma }\Big)^{8}}{1-\Big(\frac{n-G_{T}}{\sigma }\Big)^{16}}.
\end{equation}%
The parameters $\sigma $ and $\lambda $ are chosen so as to achieve the
correct ring statistics while at the same time making the function $\mathcal{%
N}_{n}(\{\mathbf{r}_{i}\})$ smooth enough to avoid problems in the
integration of the equations of motion. In our calculation we have chosen $%
\sigma =0.3$ and $\lambda =150$.

The initial proton-disordered configurations of ice were generated using the
Montecarlo procedure described in ref. \cite{buch98}, which allows
supercells with zero dipole moment to be produced. 
We employ the non-polarizable TIP5P force field \cite{mahoney} and treat the
long-range electrostatics exactly by the particle mesh Ewald algorithm.
Although this model fails in reproducing the thermodynamics of water at high
temperature ($>$500 K) \cite{lisal02}, the hydrogen bond dynamics \cite%
{paolowater}, the self-diffusion coefficient and the density anomaly are
well reproduced in the range of temperature of interest to this study ($\sim 
$270 K) \cite{mahoney,lisal02,mahoney2}. The melting of ice I$_{h}$ was
simulated in the constant pressure (NPT) ensemble in models consisting of
576 and 360 water molecules. No size effects on the results were observed
while reducing the size of the system. The results reported below refer to
simulations of 576 molecules system. In these metadynamics runs the
time-dependent potential (eq.\ref{h2}) is made up of Gaussians $0.5$
kcal/mol high, placed every $\Delta t=1$ ps. The widths $\delta s$ of the
Gaussians in the space of the collective variables are 10 and 20 for five
and six-membered rings and 24 kcal/mol for the potential energy. This choice
of parameters leads to an accuracy of $\sim 1.5$ kcal/mol \cite{metadyn04}.
\begin{figure}[t]
\begin{center}
\epsfxsize=8. truecm
\centerline{\epsffile{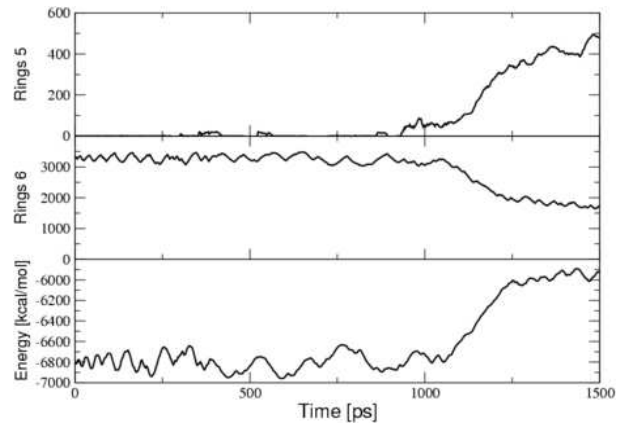}}
\caption{
Evolution of the collective variables during a metadynamics
trajectory for the 576 molecules model. Note that the metadynamics time does
not correspond to real time.}
\label{melting}
\end{center}
\end{figure}

The evolution of the collective variables during the melting is reported in
Fig.\ref{melting}. Before the phase transition takes place, a few events of
formation and recombination of 5-membered rings occur. These events
correspond to the formation of \textquotedblleft 5+7\textquotedblright\
topological defects, which recombine after a short time. This defect was
recently discovered by Grishina and Buch \cite{grishina04} and its five
different conformations arising from the possible proton arrangements
analyzed. 
%

In a separate metadynamics run we computed the free energy of the defect
using as collective variables the orientation of the O$\ldots $O bond with
respect to two cartesian axes, $x$ and $z$. Simulations were performed in
the NVT ensemble at two different temperatures, namely 120 and 270 K. The
time-dependent potential is made up of the sum of Gaussians 0.167 kcal/mol
high, placed every 1 ps, with $\delta s=0.03$ in both the dimensions of the
collective variables space.

\begin{figure}
\begin{center}
\epsfxsize= 8.5 truecm
\epsffile{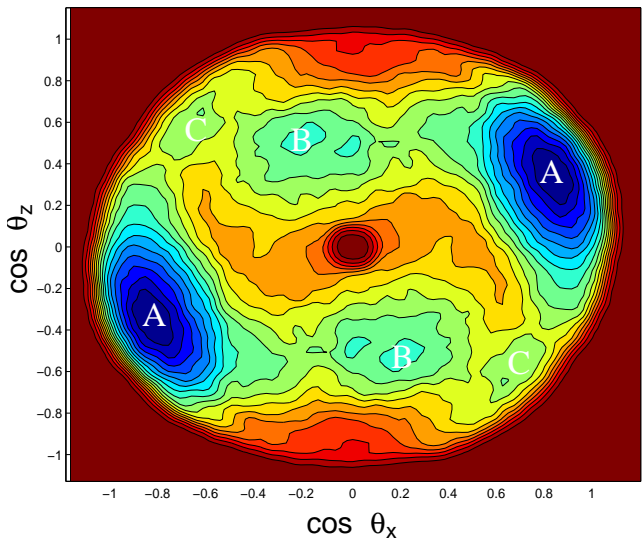}
\caption{
FES of ice I$_h$ with respect to the orientation of the bond, which
is being rotated to form the ``5+7'' defect at 120 K. }
\label{fes}
\end{center}
\end{figure}

The FES obtained by metadynamics for the topological defect formation is
represented in Fig. \ref{fes}. The FES displays a central symmetry, due to
the fact that interchanging the two water molecules used to define the space
of the collective variable has no effect on the energy of the system.
Besides the two deep basins (labeled A in Fig. \ref{fes}) that correspond to
the ideal crystal configuration, two inequivalent metastable structures have
been identified (B and C in Fig. \ref{fes}). They both correspond to 5+7
defects, but the formation of five and seven-membered rings is achieved by
different rotations of the pair of H$_{2}$O molecules, leading to structures
of varying stability, in agreement with the analysis of ref. \cite%
{grishina04}. The free energy of the most stable defect structure (B) is 6.9
kcal/mol higher than that of the ideal crystal, which we assume as the
reference zero value. The free energy barriers for the defect formation and
recombination are estimated to be 8.9 and 2.0 kcal/mol, respectively, which,
assuming a characteristic frequency for the reaction coordinates of $\sim 5$
THz and using the classical transition state theory, gives an estimated
lifetime of the defect of $\sim 0.5$ ns at 120 K. The second defect
structure (C) explored during the metadynamics run is less stable (F=8.4
kcal/mol) and can either recombine or transform into structure B through a
barrier of 0.9 kcal/mol. Raising the temperature to 270 K, the height of the
free energy minimum corresponding to structure B and its formation and
recombination barriers remain unchanged, while the recombination barrier for
structure C goes to zero. 
These free energy calculations extend the results of \cite{grishina04} to
finite temperature and confirm the relevance of these topological defects
also at the melting temperature.

\begin{figure}
\begin{center}
\epsfxsize= 8. truecm
\epsffile{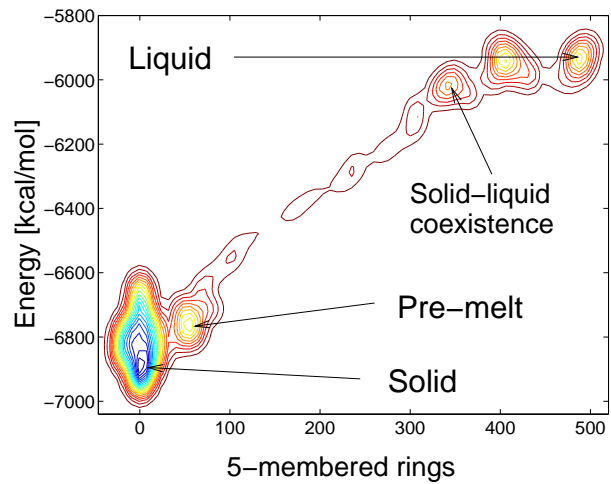}
\caption{
Two-dimensional projection of the FES into the space of the two
collective variables energy and 5-membered rings. The metadynamics run has
been interrupted before the basin corresponding to the liquid state was
explored.}
\label{melt_fes}
\end{center}
\end{figure}

\begin{figure}
\begin{center}
\epsfxsize= 8. truecm
\epsffile{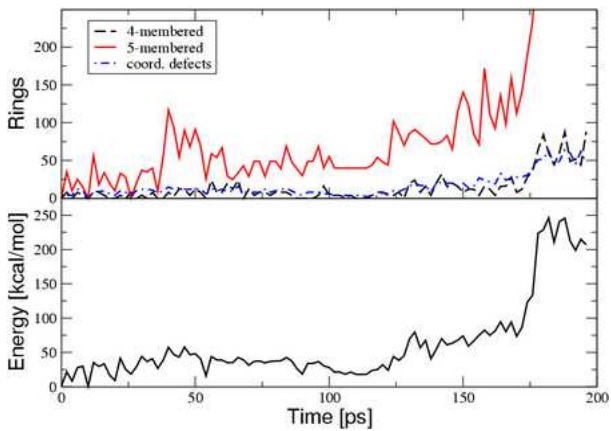}
\caption{
Defect statistics and energy of the inherent structures visited by
the system in the pre-melt region. The zero of this graph corresponds to the
beginning of the melting transition in a metadynamics run of a model with
576 water molecules.}
\label{inhst}
\end{center}
\end{figure}

\begin{figure}
\begin{center}
\epsfxsize= 8. truecm
\epsffile{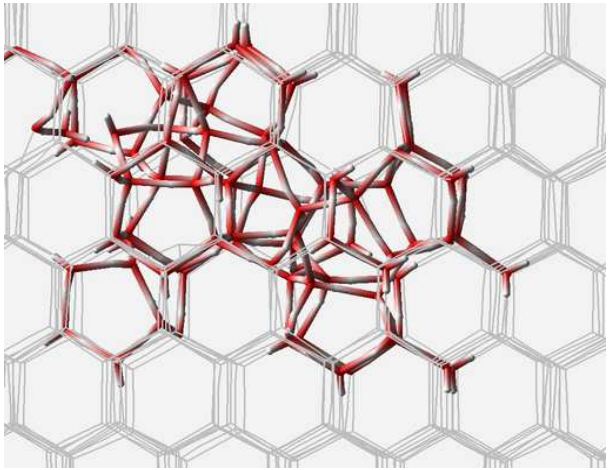}
\caption{
A cluster of topological defects obtained during the inherent
structure analysis of the pre-melting phase. H$_{2}$O molecules forming
either 4 or 5-membered rings are represented by colored sticks. Grey lines
represent the ice I$_{h}$ structure embedding the cluster of defects.}
\label{snap}
\end{center}
\end{figure}

We found that the \textquotedblleft 5+7\textquotedblright\ defects play an
even greater role, since they are responsible for the shallow minimum in the
FES indicated as pre-melt in Fig. \ref{melt_fes}. From metadynamics the
transition barrier to this local minimum can be estimated at roughly 12
kcal/mol. We analyzed the nature of this local minimum by performing an
inherent structure analysis \cite{weber82} of the metadynamics trajectory,
quenching to zero K in 50 ps frames of the MD trajectory taken every 2 ps.
As shown in Fig. \ref{inhst}, the inherent structures display a relevant
quantity of 5-membered rings and a smaller number of 4-membered rings and
coordination defects. In fact these structures correspond to a condensation
of topological defects involving about 50 H$_{2}$O molecules in an otherwise
perfect ice I$_{h}$ lattice. One typical defect cluster thus obtained is
shown in Fig. \ref{snap}. The energy of the particles, either belonging to
smaller rings or under-coordinated, ranges from 0.60 to 0.95 kcal/mol
relative to the energy of ice I$_{h}$.

These defective structures were then embedded in a crystalline ice I$_{h}$
supercell containing 4608 molecules and brought to 270 K, in order to
observe their stability. Several MD simulations were performed in the NPT
ensemble with different random initial velocities. Remarkably, the average
lifetime of the cluster of topological defects is 0.4$\pm $0.1 ns. This
relatively stable accumulation of defects in a restricted region of the
crystal is a nucleus for further disorder and melting as the number of
coordination defects and smaller rings grows. 
In Fig. \ref{inhst} two distinct regimes in the pre-melting inherent
structures can be observed. When the system is dragged out of the
free-energy basin corresponding to the cluster of topological defects, a
relevant number of small rings and coordination defects appears and the
energy suddenly increases. Roughly speaking this signals the watershed
between configurations belonging to the basin of attraction of ice I$_{h}$
and to that of the liquid.

Since the topological defects cannot migrate, the mobility of the defective
droplet is related only to defect formation and recombination at the
interface with the crystal, and no relevant motion of its center of mass was
observed. We computed the momentum of inertia of the defective region and
found that the inertia tensor has two eigenvalues I$_{1}$ and I$_{2}$ of
similar size and a smaller one I$_{3}$, with an asphericity ratio $\frac{%
(I_{1}+I_{2})/2-I_{3}}{(I_{1}+I_{2})/2+I_{3}}\simeq 0.4$. This indicates
that the cluster of defects has a somewhat elongated shape. The analysis of
the eigenvectors shows that the defective region is roughly aligned along
the ($3{\overline{3}}1$) crystallographic direction. 
Another feature of the FES that is worth commenting upon is the shallow
basin that appears just before the larger liquid basin and corresponds to a
liquid-solid interface.

In summary, our simulations reveal that topological defects where 5 and
7-membered rings are formed play a crucial role in the bulk melting of ice.
It is worthy of note that in their landmark simulation of ice nucleation
Matsumoto \textit{et al.} \cite{matsu02} found that clusters with a similar
structure to ours provide the nuclei for crystallization. The elongated
shape of these extended defects and their preferential orientation along a
given crystallographic direction might facilitate their experimental
detection. The recent report that the radiation-induced amorphization
process in silicon proceeds via the formation and aggregation of
\textquotedblleft 5+7\textquotedblright\ defects \cite{marques03} leads us
to believe that this is a general feature of disordering processes in
tetrahedral networks.

\acknowledgments{
We thank A. Laio for many useful suggestions, including the use of the potential energy
as an order parameter, and for reading the manuscript. 
}


\begin{thebibliography}{50}

\bibitem{mishima84}
O. Mishima, L.~D. Calvert, and E. Whalley, Nature (London) {\bf 310},  393  (1984).

\bibitem{fecht92}
H.~J. Fecht, Nature {\bf 356},  133  (1992).

\bibitem{laio02}
A. Laio and M. Parrinello, Procs. Nat. Acad. Sci. USA {\bf 99},  20  (2002).

\bibitem{marcella}
M. Iannuzzi, A. Laio, and M. Parrinello, Phys. Rev. Lett. {\bf 90},  238302
  (2003).

\bibitem{micheletti}
C. Micheletti, A. Laio, and M. Parrinello, Phys. Rev. Lett. {\bf 92},  170601
  (2003).

\bibitem{grishina04}
N. Grishina and V. Buch, J. Chem. Phys. {\bf 120},  5217  (2004).

\bibitem{matteo}
M. Ceccarelli, C. Danelon, A. Laio, and M. Parrinello, Biophys. J. {\bf 87},
  58  (2004).

\bibitem{iannuzzi04}
M. Iannuzzi and M. Parrinello, Phys. Rev. Lett. {\bf 93},  025901  (2004).

\bibitem{roman}
R. Martonak, A. Laio, and M. Parrinello, Phys. Rev. Lett. {\bf 90},  075503
  (2003).

\bibitem{q6}
P.~J. Steinhardt, D.~R. Nelson, and M. Ronchetti, Phys. Rev. B {\bf 28},  784
  (1983).

\bibitem{trout03}
R. Radhakrishnan and B.~L. Trout, J. Am. Chem. Soc. {\bf 125},  7743  (2003).

\bibitem{king}
S.~V. King, Nature {\bf 213},  1112  (1967).

\bibitem{yuan02}
X.~L. Yuan and A.~N. Cormack, Comp. Mat. Sci. {\bf 24},  343  (2002).

\bibitem{swfun}
The hydrogen bond function is defined as:
\begin{equation}
\nonumber f_{ij}= \frac{1-\Big(\frac{r_{ij}-C}{r_O}\Big)^{10}}{1-\Big(\frac{r_{ij}-C}{r_O}\Big)^{20}}\sum_{k}^{\rm
hydrogens}
\frac{1-\Big(\frac{r_{ik}+r_{jk}-r_{ij}}{r_H}\Big)^{8}}{1-\Big(\frac{r_{ik}+r_{jk}-r_{ij}}{r_H}\Big)^{12}}
\end{equation}
with $C=2.7$ \AA , $r_O=0.5$ \AA\ and $r_H=0.6$ \AA .
The first term in the product is a function of the O$_i$O$_j$ distance which is 1 at 2.7 \AA\ and
decays to zero at 3.5 \AA . The second term is one when the difference
O$_i$H$_k +$O$_j$H$_k -$O$_i$O$_j$ is zero and tends to zero when it exceeds 0.6 \AA .

\bibitem{paolowater}
P. Raiteri, A. Laio, and M. Parrinello, Phys. Rev. Lett. {\bf 93},  087801
  (2004).

\bibitem{buch98}
V. Buch, P. Sandler, and J. Sadlej, J. Phys. Chem. B {\bf 102},  8641  (1998).

\bibitem{mahoney}
M.~W. Mahoney and W.~L. Jorgensen, J. Chem. Phys. {\bf 112},  8910  (2000).

\bibitem{lisal02}
M. L{\'i}sal, J. Kolafa, and I. Nezbeda, J. Chem. Phys. {\bf 117},  8892
  (2002).

\bibitem{mahoney2}
M.~W. Mahoney and W.~L. Jorgensen, J. Chem. Phys. {\bf 114},  363  (2001).

\bibitem{metadyn04}
A. Laio {\it et~al.}, submitted  (2004).

\bibitem{weber82}
F.~H. Stillinger and T.~A. Weber, Phys. Rev. A {\bf 25},  978  (1982).

\bibitem{matsu02}
M. Matsumoto, S. Saito, and I. Ohmine, Nature {\bf 416},  409  (2002).

\bibitem{marques03}
L.~A. Marques {\it et~al.}, Phys. Rev. Lett. {\bf 91},  135504  (2003).

\end{thebibliography}
\end{document}